\documentclass[twocolumn,showpacs,preprintnumbers,amsmath,amssymb]{revtex4}
\usepackage{graphicx}% Include figure files
\usepackage{dcolumn}% Align table columns on decimal point
\usepackage{bm}% bold math

\begin{document}

\preprint{APS/123-QED}

\title{Phase diagram and critical properties of the frustrated Kondo necklace model
in a magnetic field}% Force line breaks with \\

\author{Takahiro Yamamoto}
\author{K{\^o}ichir{\^o} Ide}
\author{Chikara Ishii}%
\affiliation{%
Department of Physics, Faculty of Science, Science University of Tokyo,
1-3 Kagurazaka, Shinjuku-ku, Tokyo 162-8601}%

\date{
%\today
}% It is always \today, today,

\begin{abstract}
The critical properties of the frustrated Kondo necklace model with
a half saturation magnetization ($m=1/2$)
have been studied by means of an exact-diagonalization method. 
It is shown from bosonization technique that the model can be
effectively expressed as a quantum sine-Gordom model.
Thus it may show three (dimer plateau, N{\'e}el plateau and Tomonaga-Luttinger liquid)
phases due to competitions among the Ising anisotropy $\Delta$, and
the nearest- and next-nearest-neighbor exchange interactions $J_1$ and $J_2$. 
The boundary lines on the $\Delta-J_2/J_1$ phase diagram separating the three phases are
determined by the method of level spectroscopy based on the conformal field theory.
\end{abstract}

%75.10.Jm Quantized spin models
%75.30.Kz   Magnetic phase boundaries (including magnetic transitions, metamagnetism, etc.)  

\maketitle

\section{Introduction}

Intermetallic compounds containing rare-earth or 
actinide elements are known to exhibit a variety of low-temperature phases, 
including antiferromagnetism, superconductivity, and Kondo-insulator phases.
The Kondo lattice model is one of the standard models 
describing such physics of heavy electron systems.  
The low-temperature properties of the Kondo lattice model at half-filling,
describing a group of compounds called Kondo-insulators,
are governed by competition between the Kondo- and 
the Ruderman-Kittel-Kasuya-Yosida (RKKY) interactions. 
While the Kondo interaction favors the formation of
intra-atomic quantum disordered (Kondo singlet) phase, 
the RKKY interaction favors an antiferromagnetic long-range order.
In one-dimension, the Kondo singlet phase is stabilized in the presence of
Kondo interaction of any strength, causing an energy gap in
 the charge and spin excitations,
the spin gap being always smaller than the charge gap.\cite{ref:1,ref:2}

The Kondo necklace model is a simplified version of the Kondo lattice model in
which the charge degree of freedom is frozen out in order to
describe the above-mentioned competition.\cite{ref:3}
In fact, the Kondo singlet gap $\Delta_K$ always opens in
the Kondo necklace model as long as any finite Kondo interaction is present.
\cite{ref:4,ref:5,ref:6,ref:7,ref:8}
Thus the magnetization of this model starts to grow at
a finite critical magnetic field $h=\Delta_K$, 
and continuously increases with increase in the external field.    
Note that these results for the Kondo necklace model are consistent with
the magnetization curve of the Kondo lattice model.~\cite{ref:9}

However, the above situation may be drastically changed when the 
next-nearest-neighboring conduction electron spins coupled antiferromagnetically. 
In our previous paper,\cite{ref:10} we studied a magnetization process of
the Kondo necklace model with next-nearest-neighbor interaction $J_2$ which
partially frustrates the nearest-neighbor interaction $J_1$,
making use of a finite size scaling analysis of
the numerical exact-diagonalization data,

In Ref.~10
we showed that for $\lambda=J_2/J_1=1/2$, the magnetization plateaus appear at
zero ($m=0$) and a half saturation magnetization ($m=1/2$),
satisfying the necessary condition for the appearance of plateaus.~\cite{ref:11}
The origin of these plateaus for $\lambda=1/2$ can be understood as follows.
The spin configuration in the plateau phase at $m=0$ is interpreted as
the aggregate of intra-atomic Kondo singlets.
On the other hand, the spin configuration in the plateau phase at $m=1/2$
represents the aggregate of intrachain dimers  with broken translational symmetry,
resulting from the frustrations due to competition between $J_1$ and $J_2$.
However, the $m=1/2$ plateau may be completely broken by
quantum fluctuations with decreasing $J_2$
In fact, the $m=1/2$ plateau does not appear at a magnetization curve of
the Kondo necklace model without next-nearest-neighbor interaction. 
This suggests that a quantum phase transition between the plateau and nonplateau
phases may occur in $0<\lambda<1/2$.

In this paper, we are going to investigate critical properties of
frustrated Kondo necklace model 
\begin{eqnarray} 
{\mathcal H}_{\rm FKN}&=&\sum_{j=1}^{L}\left\{
\left({\mbox{\boldmath $\sigma$}}_j\cdot
{\mbox{\boldmath $\sigma$}}_{j+1}\right)_\Delta
+\lambda\left({\mbox{\boldmath $\sigma$}}_j\cdot
{\mbox{\boldmath $\sigma$}}_{j+2}\right)_\Delta
\right\}\nonumber\\
& &+J_K\sum_{j=1}^{L}{\mbox{\boldmath $\sigma$}}_j\cdot
{\mbox{\boldmath $S$}}_j,
\label{eq:knm}
\end{eqnarray}
removing the restriction $\lambda=1/2$ imposed in our previous work.\cite{ref:10} 
Here $\left({\mbox{\boldmath $\sigma$}}_j\cdot
{\mbox{\boldmath $\sigma$}}_{j+\delta}\right)_\Delta$
with $\delta=1,2$ denotes the anisotropic spin interaction of the XXZ type,
\begin{eqnarray}
\left({\mbox{\boldmath $\sigma$}}_j\cdot
{\mbox{\boldmath $\sigma$}}_{j+\delta}\right)_\Delta
=\sigma_j^x\sigma_{j+\delta}^x+\sigma_j^y\sigma_{j+\delta}^y
+\Delta \sigma_j^z\sigma_{j+\delta}^z
\end{eqnarray}
with Ising anisotropy $\Delta$,
${\mbox{\boldmath $\sigma$}}_j$ are intra-chain spins with $\sigma=1/2$,
and ${\mbox{\boldmath $S$}}_j$ are innercore spins with $S=1/2$,
hanging from the $\sigma$-spin chain.
We choose the nearest-neighbor exchange $J_1$ as the unit of energy,
and the Kondo exchange coupling $J_K$ is chosen to be antiferromagnetic ($J_K>0$).
In the case of the ferromagnetic Kondo coupling ($J_K<0$),
this model (\ref{eq:knm}) has a connection with the Haldane problem.\cite{ref:12}

We are also interested in whether or not the above-mentioned non-plateau phase
belongs to the universality class of Tomonaga-Luttinger (TL) liquid.
This problem is nontrivial in the case of the Kondo necklace model,
although it is well known that many one-dimensional quantum spin systems
exhibit the TL liquid phase with central charge $c=1$ in the conformal field theory.

The paper is organized as follows.
In \S \ref{sec:2}, we map model (\ref{eq:knm}) onto
the quantum sine-Gordon model by means of bosonization technique,
limiting ourselves to the Hilbert space with the half-magnetized states.
In \S \ref{sec:3A}, the phase diagram in the plane of Ising anisotropy $\Delta$
and the frustration parameter $\lambda$ will be constructed,
making use of the method of level spectroscopy,
on the basis of the assumption that our model (\ref{eq:knm}) is
described as the quantum sine-Gordon model.
In \S \ref{sec:3B}, the validity of this assumption will be
checked by calculating scaling dimensions and the central charge.
It will be shown that the central charge $c$ is very close to $1$. 
The final section (\S \ref{sec:4}) is devoted to a summary and discussion.

\section{\label{sec:2}Bosonization and Effective Hamiltonian}
Let us consider the effective Hamiltonian
that describes our model in the $m=1/2$ plateau phase.
We showed in our previous paper~\cite{ref:10} that
inner core spins of the Kondo necklace model with $\lambda=1/2$ in
the $m=1/2$ plateau phase are almost ferromagnetically aligned
along the external field ($z$-axis).
This enables us to replace the inner core spin operators 
${\mbox{\boldmath $S$}}_j$ with an average $\langle S_j^z\rangle=1/2$,
leading to the effective Hamiltonian of the following form:
\begin{eqnarray}
{\mathcal H}_{\rm eff}&=&\sum_{j=1}^{L}\left\{
\left({\mbox{\boldmath $\sigma$}}_j\cdot
{\mbox{\boldmath $\sigma$}}_{j+1}\right)_\Delta
+\lambda\left({\mbox{\boldmath $\sigma$}}_j\cdot
{\mbox{\boldmath $\sigma$}}_{j+2}\right)_\Delta
\right\}\nonumber\\
& &-h_{\rm eff}\sum_{j=1}^{L}\sigma_j^z
\label{eq:eff}.
\end{eqnarray}
Here $h_{\rm eff}=(h-J_{K}/2)$ is the effective magnetic field including
the effect of Kondo coupling.
Since the total magnetization of the $\sigma$ spins is vanishing
($J_{K}/2$ ($h_{\rm eff}\approx 0$), the system can be effectively regarded as
a $\sigma=1/2$ XXZ chain with next-nearest-neighbor interaction in the zero field.
The ground state properties of the model have been extensively studied by
many authors by means of various methods .~\cite{ref:13,ref:14,ref:15,ref:16}
It is known that the model is transformed into the quantum sine-Gordon model
\begin{eqnarray}
{\mathcal H}_{\rm SG}&=&\int\!\!dx \Big[
\frac{\pi v_F K}{2}\Pi^2(x)+\frac{v_s}{2\pi K}\left(\partial_x\phi(x)\right)^2
\nonumber\\
& &-\frac{2g}{\left(2\pi\alpha\right)^2}\cos\left(\sqrt{8}\phi(x)\right)
\Big],
\label{eq:sgm}
\end{eqnarray}
in terms of the Jordan-Wigner transformation and bosonization technique.
Here, $\Pi(x)$ is the momentum density conjugate to the bosonic field $\phi(x)$,
satisfying the commutation relation $\left[\phi(x), \Pi(x')\right]=i\delta(x-x')$,
and $\alpha$ is the lattice constant.
The spin wave velocity $v_s$, the Luttinger liquid parameter $K$,
and the umklapp scattering amplitude $g$ are respectively given as
\begin{eqnarray}
v_F=2\sqrt{AC},\quad K=\frac{1}{2\pi}\sqrt{\frac{C}{A}},\quad g=2\pi^2\lambda^2D,
\end{eqnarray}
where the coefficients $A$, $C$, and $D$ are given by
\begin{eqnarray}
A&=&\frac{\alpha}{8\pi}\left(1+\frac{3\Delta}{\pi}+
\frac{(6+\Delta)\lambda}{\pi}\right),\nonumber\\
C&=&2\pi\alpha\left(1-\frac{\Delta}{\pi}-
\frac{(2-\Delta)\lambda}{\pi}\right),\nonumber\\
D&=&\frac{1}{2\alpha}\left(\Delta-(2+\Delta)\lambda\right).
\end{eqnarray}
Thus the critical properties of our system with $m=1/2$ are well described in
terms of the quantum sine-Gordon model, although the above expressions for
$A$, $C$, and $D$ are valid only when $\Delta$, $\lambda$ and $h_{\rm eff}$ are small.
It is well known that the quantum sine-Gordon model exhibits two gapful
(corresponding to dimer plateau and N{\'e}el plateau)
phases and a gapless (TL liquid) phases, depending on the value of $K$.
These results are derived from the flow diagram of renormalization group equations
derived from the scaling of the cut-off $\alpha\to{\rm e}^{{\rm d}l}\alpha$ where $l=\ln L$.

Thus the frustrated Kondo necklace model belongs to
a universality class of TL liquid
if the nonlinear term is renormalized to zero as $L\to\infty$.
In the TL phase the $\sigma$-spin excitation is gapless,
and the $\sigma$-spin correlation functions algebraically decay as
\begin{eqnarray}
(-1)^r\left\langle\sigma_0^z\sigma_r^z\right\rangle
-\left\langle\sigma^z\right\rangle^2\sim r^{-K},\\
(-1)^r\left\langle\sigma_0^+\sigma_r^-\right\rangle\sim r^{-1/K}.
\end{eqnarray}
On the other hand, the dimer plateau phase is characterized by
the $\sigma$-spin excitation gap,
the exponential decay of the $\sigma$-spin correlation, and dimer long-rang order.
The N{\' e}el plateau phase is characterized by the Ising gap for $\sigma$-spin and
antiferromagnetic long-rang order (N{\' e}el order).
Also, it is known that the dimer-TL and the N{\'e}el-TL transitions are of
the Berezinskii-Kosterlitz-Thouless (BKT) type,
and the dimer-N{\'e}el transition is of the Gaussian type.

\section{\label{sec:3}Numerical Results}     
\subsection{\label{sec:3A}Phase diagram}
In this section, we are going to determine the boundaries between
TL, dimer, and N{\' e}el phases on the plane of Ising anisotropy $\Delta$ and
the frustration parameter $\lambda$, making use of the method of level spectroscopy.
This method is known to be powerful in the determination of the BKT transition points,
which is difficult by the standard finite size scaling analysis due to logarithmic corrections.
\cite{ref:16,ref:17,ref:18}

Since the Hamiltonian (\ref{eq:knm}) is invariant under spin rotation around the $z$-axis,
translation $j\to j+1$, and space inversion $j\to L-j+1$,
the eigenvalues of total magnetization $M$, wave number $k=2\pi n/L$,
and parity ${\mathcal P}=\pm 1$ are good quantum numbers to
label the eigenvalues and eigenvectors of ${\mathcal H}_{FKN}$.

As expected from bosonization,
the ground state with $m=M/L=1/2$ in all three phases (i.e., TL, dimer and N{\'e}el phases)
is shown to be labeled by $(M,k,{\mathcal P})=(L/2,0,1)$
for the lattice with $L=4\times{\rm (integer)}$
and $(M,k,{\mathcal P})=(L/2,\pi,-1)$ for the lattice with $L=4\times{\rm (integer)}+2$
On the other hand, the lowest excited state in
TL, dimer, and N{\'e}el phases are labeled by
$(M,k,{\mathcal P})=(L/2\pm 1,\pi,-1)$, $(L/2,\pi,1)$, and
$(L/2,\pi,-1)$ for $L=4\times{\rm (integer)}$ and
$(M,k,{\mathcal P})=(L/2\pm 1,0,1)$, $(L/2,0,-1)$, and
$(L/2,0,1)$ for the lattice with $L=4\times{\rm (integer)}+2$, respectively.
Hereafter, we call each of the excitations in TL, dimer, and N{\'e}el phases
{\itshape spin-wave type}, {\itshape dimer}, and {\itshape N{\'e}el} {\itshape excitaions}. 
Table \ref{table:1} summaries the eigenvalues of these symmetry operators in
the ground state and three (spin-wave type, dimer and N{\'e}el excitations)
excitations in the frustrated Kondo necklace model with $m=1/2$. 
In the case of $L=4\times$(integer), the (finite size) critical points are
determined from the crossing of the minimum excitation energies of
the above three types of excitations
\begin{eqnarray}
\Delta E_{\rm sw}&=&\frac{1}{2}\Big\{E(M=L/2+1,k=\pi,P=-1)\nonumber\\
& & +E(M=L/2-1,k=\pi,P=-1)\nonumber\\
& & -2E(M=L/2,k=0,P=1)\Big\},\label{eq:sw}\\
\Delta E_{\rm dimer}\quad &=&E(M=L/2,k=\pi,P=1)\nonumber\\
& &-E(M=L/2,k=0,P=1),\\
\Delta E_{\rm N{\text{\'e}}el}\quad&=&E(M=L/2,k=\pi,P=-1)\nonumber\\
& &-E(M=L/2,k=0,P=1).
\end{eqnarray}  

%%%%%% Table.1: Symmetry classification
\begin{table}
\caption{Symmetry classification of the ground state and three 
(spin-wave type, dimer and N{\'e}el excitations) excitations in
the Kondo necklace model with $m=1/2$ and $L=4\times{\rm (integer)}$.
The numbers in round brackets are quantum numbers in case of $L=4\times{\rm integer}+2$.}
\label{table:1}
\begin{ruledtabular}
\begin{tabular}{c c c c}
                        &             & Quantum Numbers &                \\
                        & $M$         & $k$             & ${\mathcal P}$ \\ \hline
ground state	        & $L/2$ 	  & $0$ $(\pi)$     & $+1$ $(-1)$    \\
spin-wave excited state & $L/2\pm 1$  & $\pi$ $(0)$     & $-1$ $(+1)$    \\
dimer excited state	    & $L/2$ 	  & $\pi$ $(0)$     & $+1$ $(-1)$    \\
N{\'e}el excited state	& $L/2$ 	  & $\pi$ $(0)$     & $-1$ $(+1)$    \\
\end{tabular}
\end{ruledtabular}
\end{table}

Note that these minimum excitation energies correspond to
the scaling dimensions $x_{\rm sw}$, $x_{\rm dimer}$, and $x_{\rm N{\text{\'e}}el}$ by
\begin{eqnarray}
x_i=\frac{L}{2\pi v_s}\Delta E_i\quad(i={\rm sw,\ dimer\ and\ N{\text{\'e}}el}),
\end{eqnarray}
where $v_s$ is the spin wave velocity defined as 
\begin{eqnarray}
v_s=\lim_{L\to\infty}\frac{L}{2\pi}\Big\{&E&(M=L/2,k=2\pi/L)\nonumber\\
& &-E(M=L/2,k=0)\Big\}.
\end{eqnarray}
We defined the gap in the spin-wave type excitation $\Delta E_{\rm sw}$ as
in eq.(\ref{eq:sw}) so that the Zeeman energy due to the external magnetic field
\begin{eqnarray}
{\mathcal H}_{\rm Zeeman}=-h\sum_{j=1}^{L}\left(\sigma_j^z+S_j^z\right)
\end{eqnarray}
is canceled out.

%%%%% Fig.1: crossing point

\begin{figure}[b]
  \begin{center}
    \includegraphics[keepaspectratio=true,height=60mm]{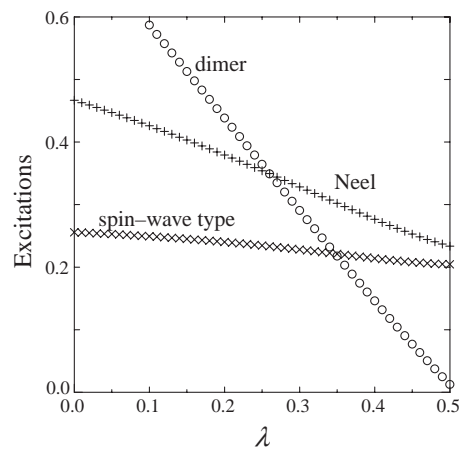}
  \end{center}
  \caption{The dimer, N{\'e}el, and spin-wave type excitation energies as
  functions of the frustration parameter $\lambda$.
  The crossing point of dimer and spin-wave type excitaion energies represents
  the BKT transition point for $L=8$ and $\Delta=1.0$.}
\label{fig:1}
\end{figure}

Let us consider the TL-dimer critical point for $\Delta=J_K=1.0$ in order to
illustrate how our method of level-crossing works. 
In Fig.\ref{fig:1}, spin-wave, dimer and N{\'e}el excitation gaps for $L=8$ are
shown as functions of the parameter $\lambda$.
We see from Fig.\ref{fig:1} that the spin-wave type excitation is the lowest for
$\lambda<0.35105$ and so is the dimer excitation for $\lambda>0.35105$.
On the other hand, the N{\'e}el excitation is always higher than spin-wave type
and/or dimer excitations.

\begin{figure}[htbp] 
  \begin{center}
    \includegraphics[keepaspectratio=true,height=60mm]{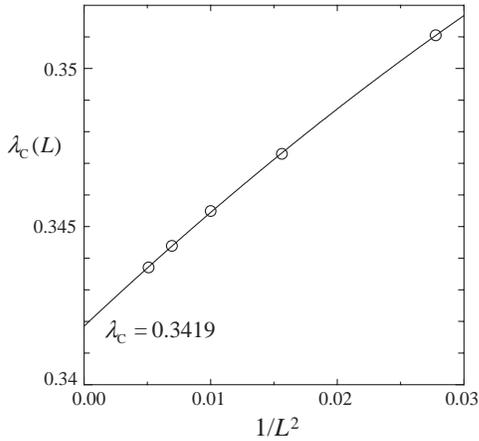}
  \end{center}
\caption{The crossing points of dimer and spin-wave type excitation energies for 
various sizes ($L=6,8,10,12$ and $14$) and $\Delta=1.0$. 
In the thermodynamic limit, 
the BKT transition point is obtained as $\lambda_c=0.3419\pm0.0001$, 
assuming an extrapolation formula $\lambda_c(L)=\lambda_c+A/L^2+B/L^4$,
assuming an extrapolation formula $\lambda_c(L)=\lambda_c+A/L^2+B/L^4$.}
\label{fig:2}
\end{figure}

\begin{figure}[t]  
  \begin{center}
    \includegraphics[keepaspectratio=true,height=60mm]{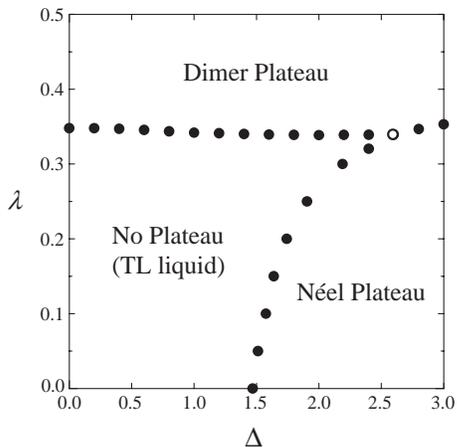}
  \end{center}
\caption{Phase diagram on the $\Delta-\lambda$ plane of 
the frustrated Kondo necklace model with $m=1/2$ for $J_K=1.0$.
The open circle denotes a multicritical point between the dimer,
N{\' e}el and TL phases.}
\label{fig:3}
\end{figure}

We have estimated the values of $\lambda$ at the crossing point of spin-wave type and
dimer excitations for various sizes $L=6,8,10,12$ and $14$.
These are extrapolated to the thermodynamic limit $L\to\infty$,
assuming a formula $\lambda_c(L)=\lambda_c+A/L^2+B/L^4$.
Thus the TL-dimer critical point $\lambda_c=0.3419\pm 0.0001$ (See Fig.\ref{fig:2})
is obtained in the limit $L\to\infty$.
We have performed similar procedures for various $\Delta$'s
to construct the phase diagram on $\Delta-\lambda$ plane, as shown in Fig.\ref{fig:3}.
We assumed that $m=1/2$ and  $J_K=J_1$ in these calculations.

%%%%% Fig2: extraporation of crossing point

%%%%% Fig3: Phase diagram

\subsection{\label{sec:3B}Consistency Check}
\begin{figure}[t]
  \begin{center}
    \includegraphics[keepaspectratio=true,height=60mm]{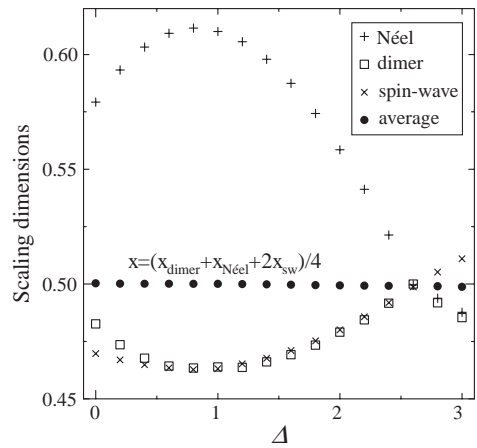}
  \end{center}
\caption{The average scaling dimension 
$x=(x_{\rm dimer}+x_{\rm N{\text{\'e}}el}+2x_{\rm sw})/2$ along 
the TL-dimer and dimer-N{\rm e}el boundaries in Fig.{\ref{fig:3}}.}
\label{fig:4}
\end{figure}

In \S \ref{sec:2}, we assumed that the effective Hamiltonian of our model at 
$m=1/2$ is described by the quantum sine-Gordon model.
Our analysis, making use of the method of level spectroscopy,
is based on this assumption, 
according to which the scaling dimension without logarithmic corrections,
\begin{eqnarray} 
x=\frac{x_{\rm dimer}+x_{\rm N{\text{\'e}}el}+2x_{\rm sw}}{4}
\label{eq:ave_x}
\end{eqnarray}
should tend to $1/2$ on the BKT critical line in the thermodynamic limit $L\to\infty$.
We confirmed this relation $x\approx 1/2$ on the TL-dimer abd dimer-N{\' e}el 
transition lines, as clearly shown in Fig.{\ref{fig:4}}.
Furthermore, Table.\ref{table:2} indicates that the relations between the scaling dimensions
\begin{eqnarray}
x_{\rm N{\text{\'e}}el}=x_{\rm dimer},
\label{eq:scal_1}\\
x_{\rm sw} x_{\rm dimer}=1/4
\label{eq:scal_2}
\end{eqnarray}
are satisfied on the dimer-N{\'e}el (Gaussian) transition points.

%%%%% Fig.4: average critical demension 

\begin{figure}[h]
  \begin{center}
    \includegraphics[keepaspectratio=true,height=60mm]{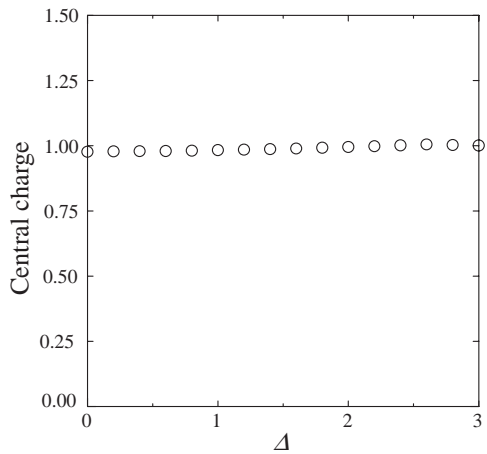}
  \end{center}
\caption{The central charge $c$ on the TL-dimer and N{\'e}el-dimer critical lines,
yielding the result $c=1$ within errors of $0.03$\%.}
\label{fig:5}
\end{figure}

We have also estimated the central charge $c$ along the TL-dimer and
dimer-N{\'e}el critical lines from the finite size correction to
the ground-state energy
\begin{eqnarray}
E_0\approx \epsilon_0 L-\frac{\pi v_s}{6L}c,
\label{eq:fsc}
\end{eqnarray}
where $\epsilon_0$ is the ground-state energy per site (including two spins)
in the thermodynamic limit.
As shown in Fig.\ref{fig:5}, we obtained the value $c\approx 1$ with small errors.
These results provide sufficient evidence that our model can be described
by the quantum sine-Gordon model.

%%%%% Fig.5: central charge

\begin{table}[t]
\caption{The relations between the scaling dimension 
[Eqs.~(\ref{eq:scal_1}) and (\ref{eq:scal_2})].}
\label{table:2}
\begin{ruledtabular}
\begin{tabular}{@{\hspace{\tabcolsep}\extracolsep{\fill}}ccc}
$\Delta$   & $x_{\rm N{\text{\'e}}el}/x_{\rm dimer}$ & $x_{\rm sw}(x_{\rm dimer}+x_{\rm N{\text{\'e}}el})/2$ \\ \hline
2.6	       & 0.999 	                    & 0.2492 \\
2.8        & 1.004                        & 0.2490 \\
3.0  	   & 1.005 	                    & 0.2486 \\
\end{tabular}
\end{ruledtabular}
\end{table}
  
\section{\label{sec:4}Conclusion}
In this paper, we have studied critical properties of 
the frustrated Kondo necklace model with a half-saturation magnetization ($m=1/2$),
making use of the method of level spectroscopy. 
The phase diagram on the plane of the Ising anisotropy $\Delta$
and the frustration parameter $\lambda$ is constructed
exhibiting two plateau (dimer and N{\'e}el) and a TL phases.
We also confirmed that the values of scaling dimension without 
logarithmic corrections [Eq.~(\ref{eq:ave_x})] are close to $1/2$ at
the TL-dimer (BKT) and dimer-N{\'e}el (Gaussian) transition points.
The relations between the scaling dimensions 
[Eqs.~(\ref{eq:scal_1}) and (\ref{eq:scal_2})] are satisfied as
the Gaussian transition points. 
Furthermore, the central charge $c\simeq 1$ is obtained along the BKT and
Gaussian transition lines, by estimating the finite-size correction for
the ground state energy [Eq.~(\ref{eq:fsc})].   
Thus we can conclude that the present system belongs to
the same universality class as the quantum sine-Gordon model.

%===<< Section Acknowledgements >>===%
\begin{acknowledgments}
We would like to thank Masaaki Nakamura for valuable advice on
the method of level spectroscopy,
and T. Nikuni for valuable comments and for a critical reading of the manuscript.
We have learned a lot about the exact-diagonalization method based on
Lanczos algorithm from the program TITPACK version 2 developed by
H. Nishimori at Tokyo Institute of Technology.
The work was supported in part by Frontier Research Center for Computational Science
(FRCCS) of Science University of Tokyo.
\end{acknowledgments}


\begin{thebibliography}{100}

\bibitem{ref:1} H. Tsunetsugu, Y. Hatsugai, K. Ueda and M. Sigrist,
Phys. Rev. B {\bf 46}, 3175 (1992).

\bibitem{ref:2} N. Shibata, T. Nishino, K. Ueda and C. Ishii,
Phys. Rev. B {\bf 53}, R 8828 (1996).

\bibitem{ref:3} S.~ Doniach,
Physica {\bf B 91}, 231 (1977).

\bibitem{ref:4} G.M. Zhang, Q. Gu and L. Yu,
Phys. Rev. B {\bf 62}, 69 (2000).

\bibitem{ref:5} R.T. Scalettar, D. J. Scalapino and R. L. Sugar,
Phys. Rev. B {\bf 31}, 7316 (1985).

\bibitem{ref:6} S. Moukouri, L. G. Caron and C. Bourbonnais and L. Hubert,
Phys. Rev. B {\bf 51}, 15920 (1994).

\bibitem{ref:7} H. Otsuka and T. Nishino,
Phys. Rev. B {\bf 52}, 15066 (1995).

\bibitem{ref:8} Y. Chen, Q. Yuan, H. Chen and Y. Zhang,
Phys. Lett. A {\bf 245}, 167 (1998).

\bibitem{ref:9} S.~Watanabe,
J. Phys. Soc. Jpn. {\bf 69}, 2947 (2000).

\bibitem{ref:10} T.~Yamamoto, M.~Asano and C.~Ishii,
J. Phys. Soc. Jpn. {\bf 122}, 345 (2001).

\bibitem{ref:11} M. Oshikawa, M. Yamanaka and I. Affleck,
Phys.\ Rev.\ Lett {\bf 78}, 1984 (1997).

\bibitem{ref:12} J.~Igarashi, T.~Tonegawa, M.~Kaburagi and P.~Fulde,
Phys. Rev. B {\bf 51}, 5814 (1995).


\bibitem{ref:13} F.D.M.~Haldane,
Phys. Rev. B {\bf 25}, 4925 (1982).

\bibitem{ref:14} K.~Kuboki and H.~Fukuyama,
J. Phys. Soc. Jpn. {\bf 56}, 3126 (1987).

\bibitem{ref:15} T.~Tonegawa, T.~Harada and M.~Kaburagi,
J. Phys. Soc. Jpn. {\bf 61}, 4665 (1992).

\bibitem{ref:16} K.~Nomura and K.~Okamoto,
J. of Phys. A: Math. Gen {\bf 27}, 5773 (1994).

\bibitem{ref:17} K.~Okamoto,
Prog. Theor. Phys. Suppl. {\bf 145}, 113 (2002). 

\bibitem{ref:18} K.~Nomura and A.~Kitazawa,
cond-mat/0201072,
in Proc. French-Japanese Symp. Quantum Properties of Low-Dimensional Antiferromagnets
(to be published)

\end{thebibliography}
\end{document}